\documentclass[twocolumn,showpacs,preprintnumbers,amsmath,amssymb]{revtex4-1}

\usepackage{graphicx}
\usepackage{bm}% bold math

\begin{document}

\title{ Efficient algortihms for the two dimensional Ising model with a surface field}
%\author{X. T. Wu an J. Indekeu }
\author{X. T. Wu}

\affiliation{Department of Physics, Beijing Normal University,
Beijing, 100875, China}

\date{\today}% It is always \today, today,

\begin{abstract}
 Bond propagation and site propagation algorithm are extended to the two dimensional Ising model with a surface field. With these algorithms we can calculate the free energy, internal energy, specific heat, magnetization, correlation function, surface magnetization, surface susceptibility and surface correlation. To test these algorithms, we study the Ising model for wetting transition, which is solved exactly by Abraham. We can locate the transition point accurately to $10^{-8}$. We carry out the calculation of the specific heat, surface susceptibility on the lattices with the sizes are up to $200^2 \times 200$. The results show that finite jump develops in the specific heat and surface susceptibility at the transition point as the lattice size increases. On the lattice with size $320^2 \times 320$ the parallel correlation length exponent is $1.88$, while in the Abraham's exact result it is $2.0$. The perpendicular correlation length exponent on the lattice with size $160^2\times 160$  is $1.04$, where its exact value is $1.0$.
\end{abstract}


\maketitle

\section{Introduction}

As we know, the Ising model with surface field is a powerful tool to study the wetting transition \cite{abraham1980,albano1990,parry1990,abraham1993,binder1995,abraham2002,abraham2003,parry,albano2012} and the wetting
transition, as an important topic in phase transition, has been intensively studied in last thirty years.
Consider a half infinite Ising ferromagnet with positive magnetization in the bulk
in zero bulk field, a negative field at the surface may stabilize a
domain with oppositely oriented magnetization at the surface. While in the ``nonwet" state of the wall the thickness of such a wetting
layer is microscopically small (i.e., a few lattice spacings in the example of
the Ising magnet), changing the surface field, one may encounter a wetting transition, where the thickness
of the wetting layer diverges, (i.e., the interface between the coexisting phases is no longer "bound" to the
surface). This model is related to a rich variety
of physical phenomena such as wetting, capillary condensation,
thin film growth, epitaxy, interface roughening, etc. The
understanding of these phenomena is of primary importance for
many technological applications in the field of material science
in general and particularly for the development of nano and
micro-devices \cite{degennes,bonn}.

In recent years, for the two dimensional Ising model without surface field, efficient algorithms called bond propagation (BP) and site propagation (SP) algorithms are developed to calculate the free energy, internal energy, specific heat and correlation \cite{loh1,loh2,wu1}. These algorithms are very efficient and can reach very high accuracy. Using BP algorithm, the Ising model on square and triangle lattice with different shapes has been studied \cite{wu2,wu3}. The largest size of lattice can reach $8000\times 8000$. Very accurate expansions of free energy at critical point have been obtained. The corner's logarithmic corrections agree with the conformal field theory \cite{cardy} in an accuracy of $10^{-14}$. Applying these algorithms, the edge and corner terms of the internal energy and specific heat have been obtained numerically for rectangular, triangular, rhomboid, trapezoid and hexagonal shape \cite{wu2014}. The numerical result is so accurate (to the accuracy $10^{-28}$), that the exact results on these terms are conjectured.

In this paper we extend these algorithms to the two dimensional Ising model with a surface field.  With these algorithms, we can calculate the partition function, internal energy, specific heat, magnetization, correlation functions, surface magnetization, surface susceptibility and surface correlation. To test these algorithms, we study the Abraham's model, where a wetting transition takes place as the the surface filed changes \cite{abraham1980,albano1990}. Studying the intersections of the surface susceptibility of different size, we can locate the transition point accurately to $10^{-8}$. We carry out the calculation of the specific heat, surface susceptibility on the lattices, of which the parallel and perpendicular sizes up to $200^2$ and $200$ respectively. The results show finite jumps in the specific heat and surface susceptibility at the transition point. The direct calculation of the correlation length on the lattice with size $320^2 \times 320$ shows that the effective parallel and perpendicular correlation length exponent is $1.88$ and $1.04$ respectively, where the exact Abraham's result is $2.0$ and $1.0$ respectively \cite{abraham1980}. Therefore these algorithms can be expected to be a powerful tool to investigate the wetting transition.

Our paper is arranged as follows. In section 2, the algorithms are derived. In section 3, the algorithms are applied to the Ising model for wetting transition. Section 4 is a summary.

\section{Algorithms}

We consider the two dimensional Ising model with open boundaries and a surface field, which is defined by
\begin{equation}
-\beta \mathcal{H}=\sum_{<ij>}J_{ij}\sigma_i \sigma_j+\sum_{i\in \Gamma}H_{1i} \sigma_i.
\end{equation}
where $\beta=1/K_BT$ and $\sigma=\pm 1$.  The dimensionless couplings $J_{ij}=\beta\tilde{J}_{ij}$ and surface field $H_{1i}=\beta \tilde{H}_{1i}$ are arbitrary real number. $\Gamma$ represents all the sites at the surface. In our consideration the bulk filed is absent, i.e. the magnetic field on the interior spins is zero. By the way, it should be pointed that BP algorithm with a magnetic field being applied to a interior spin does not exist.

Generally we need to calculate the partition function
\begin{equation}
Z=\sum_{\{\sigma_i\}}e^{-\beta H},
\end{equation}
the internal energy
\begin{equation}
U=-\frac{\partial \ln Z}{\partial \beta}=\sum_{\{\sigma_i\}}He^{-\beta H}/Z,
\label{eq:internal}
\end{equation}
and specific heat
\begin{equation}
C=\beta^2 \frac{\partial ^2 \ln Z}{\partial \beta ^2}=\beta^2 (\sum_{\{\sigma_i\}}H^2e^{-\beta H}-U^2).
\label{eq:specific}
\end{equation}
To study the effect of the surface field, one may calculate the surface magnetization of a part of the surface
\begin{equation}
m_1=\frac{1}{L}\sum_{\{\sigma_i\}}(\sum_{j\in
\Gamma_1}\sigma_j)e^{-\beta H}/Z, \label{eq:boundarymag}
\end{equation}
where $\Gamma_1$ is the part of surface and  $L$ is length of  $\Gamma_1$; and the corresponding susceptibility
\begin{equation}
\chi_{11}=\frac{1}{L}[\sum_{\{\sigma_i\}}(\sum_{j\in
\Gamma_1}\sigma_j)^2e^{-\beta H}-m_1^2], \label{eq:boundarysuscep}
\end{equation}
The algorithms for these quantities are developed in the following. They are the extensions of BP algorithm \cite{loh1,loh2,wu2014}.

We also develop the an algorithm to calculate the magnetization of a spin including the interior spin and surface spin
\begin{equation}
m_j=<\sigma_j>=\sum_{\{\sigma_i\}}\sigma_je^{-\beta H}/Z,
\end{equation}
and the correlation function of two spins
\begin{eqnarray}
g({\textbf r}_j,{\textbf r}_k)& = & <\sigma_i\sigma_j>-<\sigma_i><\sigma_j> \nonumber \\
                              & = & \sum_{\{\sigma_i\}}\sigma_j \sigma_k e^{-\beta H}/Z-m_jm_k.
\end{eqnarray}
The multi point correlation function can also be developed. These algorithms are the extensions of SP algorithm \cite{wu1}.

\subsection{Surface-Field-Bond-propagation algorithm for internal energy, surface magnetization}

Similar to the bond propagation (BP) algorithm \cite{loh1,loh2,wu2014}, we propose a BP algorithm for the Ising model with a
surface field. We call it SFBP algorithm. The schematic of this algorithm is shown in Fig. 1.
A magnetic field is applied to the spins at the edges, which are represented by
circles with a cross. The new ingredient we introduce is SFBP series reduction.  Combining  with BP $\Delta-Y$ transformation
and its inverse \cite{loh1,loh2,wu2014}, a 2D lattice can be efficiently reduced to a small lattice. Starting from the upper left corner in Fig. 1(a), use the SFBP series reduction to convert the corner into a diagonal bond. Using the $Y-\Delta$ and $\Delta -Y$ transformations, this diagonal bond can be successively propagated diagonally down and to the right until it annihilates at an edge with open boundary conditions. Repeating these procedures turn the lattice into a small lattice with 4 spins as shown in Fig (c9). It can be calculated simply.

\begin{figure}
 \begin{center}
    \resizebox{9cm}{7cm}{\includegraphics{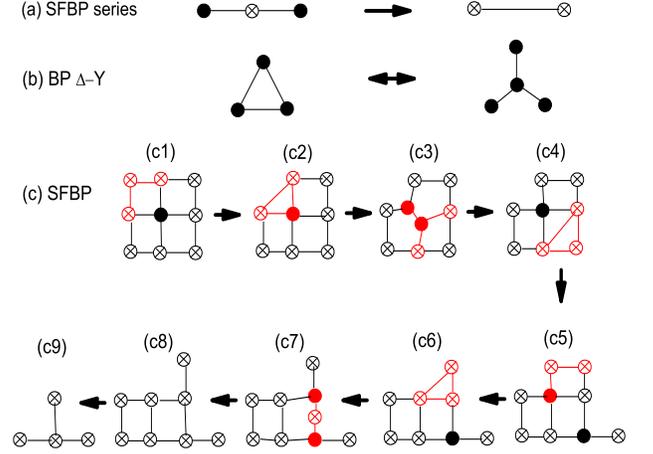}}
  \end{center}
\vskip -1cm
  \emph{
\caption{(Corlor online) The schematic of SFBP algorithm. The spins with a external field are represented by circles with a cross.
The spins without external field are represented by solid circles. (a) The SFBP series reduction transformation (see text). (b) The $\Delta -Y$ and $Y-\Delta$ transformation. Applying the SFBP series reduction to the red part in (c1) leads to (c2).
Applying $\Delta-Y$ transformation to the red part in (c2) leads to (c3). Repeated applications of the algorithm reduce the lattice to a small lattice in (c9), then we can calculate this small lattice directly.} }
\end{figure}

For the partition function, SFBP series reduction corresponds to integrating out a spin, on which a magnetic field is applied, with two neighbors, generating an effective coupling $J_{12}$ and $\delta H_1,\delta H_2, \cdots$ to make the
\begin{equation}
\sum_{\sigma_0}e^{\sigma_0(J_{10}\sigma_1+J_{20}\sigma_2+H_0\sigma_0)}\equiv
e^{\delta F+J_{12}\sigma_1\sigma_2+\delta H_1 \sigma_1+\delta H_2 \sigma_2}, \label{eq:series-pa}
\end{equation}
This equation should be valid for all $\sigma_1,\sigma_2$, then we get
\begin{eqnarray}
e^{\delta F+J_{12}\pm \delta H_1\pm \delta H_2}=2\cosh(H_0\pm J_{10}\pm J_{20})\equiv z_{\pm} , \nonumber \\
e^{\delta F-J_{12}\pm \delta H_1 \mp \delta H_2}=2\cosh(H_0\pm J_{10} \mp J_{20})\equiv y_{\pm} . \nonumber \\
\label{eq:pa-2}
\end{eqnarray}
It is convenient to use variables
$j_i=e^{-J_i},j_{ij}=e^{-J_{ij}}, \delta h_i=e^{-\delta H_i}$ and $\delta f=e^{\delta F}$. The solution of the
above equations is geiven by
\begin{eqnarray}
\delta f & = & (z_+z_-y_+y_-)^{1/4},\hskip 0.5cm \delta h_1=\delta f/(z_+y_+)^{1/2}, \nonumber \\
\delta h_2 & = & \delta f/(z_+y_-)^{1/2}, \hskip 0.5cm j_{12}=\delta f\delta h_1 \delta h_2/z_-.
\end{eqnarray}

In reference \cite{loh2}, the authors proposed the $Y-\Delta$ transformation and its inverse for the internal energy. We complete the algorithm
and proposed the algorithm, in which the BP series and $Y-\Delta$ transformations are included,  for the specific heat \cite{wu2014}.

In order to use the BP algorithm to calculate the internal energy directly, we need to calculate the following quantity
\begin{widetext}
\begin{equation}
\sum_{\{\sigma_i\}} (\sum_{ij}J'_{ij}\sigma_i\sigma_j  +
\sum_{j\in \Gamma}H'_{1j} \sigma_j+F')
\exp( F+\sum_{ij} J_{ij}\sigma_i\sigma_j+\sum_{j\in \Gamma} H_{1j}
\sigma_j).
\label{eq:bp-in}
\end{equation}
\end{widetext}
This quantity allows us to compute not only the internal energy, but also the magnetization of a surface spin, and total magnetization of a part of surface. To obtain the internal energy, we need to assign the initial values to be  $J'_{ij}=\tilde{J}_{ij}, H'_{1i}=\tilde{H}_{1i}, F=F'=0$. As the transformations are completed, the final result of $F'$ is the total internal energy. For the magnetization of a surface spin, we assign the initial values to be  $J'_{ij}=0,F=F'=0, H'_i=0$ except $H'_{1j}=1.0$ to obtain $<\sigma_j>$, where $\sigma_j$ is a spin at surface. As the transformations are completed, the final result of $F'$ is the magnetization of $j$th spin. If we want to calculate the total magnetization of a part of surface $\Gamma_1$, we can assign  $J'_{ij}=0, F=F'=0$, $H'_{1j}=1.0, \hskip 0,5cm j\in \Gamma_1$; $H'_{1j}=0$, otherwise. As the transformations are completed, the final result of $F'$ is the total magnetization of spins in the boundary $\Gamma_1$.

For the bonds on the boundary we develop SFBP series transformations which preserve the quantity in Eq. (\ref{eq:bp-in}). For the bonds in the interior of lattice, where the field terms are absent, it has been shown that this quatity can be preserved in the $\Delta-Y$ and its inverse \cite{loh2}.

The SFBP series reduction corresponds to integrating out a spin with two neighbors, generating an effective coupling
$J_{12},J'_{12},\delta H_1, \delta H_2,\delta H'_1, \delta H'_2$ to make
\begin{widetext}
\begin{eqnarray}
\sum_{\sigma_0}
 & & e^{J_{10}\sigma_0\sigma_1+J_{20}\sigma_0\sigma_2+H_0\sigma_0+H_{ex}}[(J'_{10}\sigma_0\sigma_1+J'_{20}\sigma_0\sigma_2+H'_0\sigma_0+H'_{ex})
\nonumber \\
& = & e^{\delta F+J_{12}\sigma_1\sigma_2+\delta H_1 \sigma_1+\delta H_2 \sigma_2+H_{ex}} [(\delta
 F'+J'_{12}\sigma_1\sigma_2+\delta H'_1 \sigma_1+H'_2 \sigma_2+H'_{ex})  ], \label{eq:series-in1}
\end{eqnarray}
\end{widetext}
where the terms not involving $\sigma_0$ are collected in
$H'_{ex}(\sigma_1,\sigma_2,\cdots)$.
In every step, we have $F\rightarrow F+\delta F$, $F'\rightarrow
F'+\delta F'$. Since $H'_{ex}$ contain the variables
$\sigma_1,\sigma_2,\cdots,U,C$, the coefficients before them in
the two sides of the above equation must be equal. Equating the
coefficients before $H'_{ex}$ in both sides of the above equation
yields Eq. (\ref{eq:series-pa}), which is solved above.

The rest terms in the two sides of Eq. (\ref{eq:series-in1}) should
also be equal, then we get
\begin{widetext}
\begin{eqnarray}
&&\sum_{\sigma_0}[J'_{10}\sigma_1\sigma_0+J'_{20}\sigma_2\sigma_0+H'_0\sigma_0]e^{J_{10}\sigma_1\sigma_0+J_{20}\sigma_2\sigma_0+H_0\sigma_0}
\nonumber \\
 &=&[\delta F'+J'_{12}\sigma_1\sigma_2+\delta H'_1\sigma_1+\delta H'_2\sigma_2]e^{\delta
F+J_{12}\sigma_1\sigma_2+\delta H_1 \sigma_1+\delta H_2 \sigma_2}.
\end{eqnarray}
\end{widetext}
This equation should be valid for all $\sigma_1,\sigma_2$, then we get
\begin{eqnarray}
&& e^{\delta F+J_{12}\pm \delta H_1\pm \delta H_2}(\delta F'
+J'_{12}\pm \delta H'_1\pm \delta H'_2) \nonumber  \\
& = & 2(H'_0\pm J'_{10}\pm J'_{20})\sinh(H_0\pm J_{10}\pm J_{20})
\end{eqnarray}
and
\begin{eqnarray}
& &e^{\delta F-J_{12}\pm \delta H_1\mp \delta H_2}(\delta F'
-J'_{12}\pm \delta H'_1\mp \delta H'_2) \nonumber \\
& = & 2(H'_0\pm J'_{10}\mp
J'_{20}) \sinh(H_0\pm J_{10}\mp J_{20})
\end{eqnarray}

Substituting Eqs.(\ref{eq:pa-2}) into above equations, we get
\begin{eqnarray}
\delta F' & = & \frac{1}{4}(z'_{+}+z'_{-}+y'_++y'_-), \nonumber \\
\delta H'_1 & = & \frac{1}{2}(z'_+ +y'_+)-\delta U \nonumber \\
\delta H'_2 & = & \frac{1}{2}(z'_+ +y'_-)-\delta U \nonumber \\
J'_{12}  &  = & \frac{1}{2}(z'_++z'_-)-\delta U \label{eq:bp-u}
\end{eqnarray}
where
\begin{eqnarray}
z'_{\pm} & = & (H'_0\pm J'_{10}\pm J'_{20})\tanh(H_0\pm J_{10}\pm J_{20}), \nonumber \\
y'_{\pm} & = & (H'_0\pm J'_{10}\mp J'_{20})\tanh (H_0\pm J_{10}\mp J_{20}).
\end{eqnarray}

This algorithm including SFBP series reduction, $Y-\Delta$ and its inverse allows us to compute the internal energy, magnetization of a surface spin, and total magnetization of a part of surface.

\subsection{Surface-Field-Bond-propagation algorithm for specific heat, surface susceptibility and surface correlation function}

For the specific heat Eq. (\ref{eq:specific}), the correlation between surface spins and surface susceptibility defined in (\ref{eq:boundarysuscep}),  we need to calculate the following quantity
\begin{widetext}
\begin{equation}
\sum_{\{\sigma_i\}} [(\sum_{ij}J'_{ij}\sigma_i\sigma_j +\sum_{j\in \Gamma}H'_{1j} \sigma_j+F')^2
+\sum_{ij}J''_{ij}\sigma_i\sigma_j+\sum_{j\in \Gamma}H''_{1j}
\sigma_j+F'']\exp(F+\sum_{ij}J_{ij}\sigma_i\sigma_j+\sum_{j\in
\Gamma}H'_{1j} \sigma_j).
\label{eq:bp-sp}
\end{equation}
\end{widetext}
For the specific heat, we need to assign the initial values to be $J'_{ij}=\tilde{J}_{ij}, H'_{1i}=\tilde{H}_i, J''_{ij}=F=F'=F''=0$. After some
transformations, $J''_{ij},F,F',F''$ will become nonzero. As the transformations are completed, the final result of $F,F'$ and
$F''$ are the total free energy, internal energy and specific heat of the system. For the surface susceptibility, we need to assign the initial values to be  $ H'_{1i}=1, i\in \Gamma_1; H'_{1i}=0, otherwise $ and $J'_{ij}=J''_{ij}=F=F'=F''=0$. The final results of  $F,F',F''$ are the free energy, surface magnetization and surface susceptibility respectively. We can also calculate the correlation between two surface spins $\sigma_i,\sigma_j$ just letting $J'_{kl}=J''_{kl}=F=F'=F''=0$ and $ H'_{1i}=H'_{1j}=1; H'_{1k}=0, k \neq i,j$.

The $\Delta-Y$ transformation and its inverse to preserve the quantity in the above equation has been given by Wu et. al \cite{wu2014}. Now we present the SFBP series reduction to preserve the quantity in the above equation. This corresponds to integrating out a spin with two neighbors, generating an effective coupling $J_{12},J'_{12},J''_{12}$, and $\delta H_1, \delta H_2,\delta H'_1, \delta H'_2\delta H''_1, \delta H''_2$ to make
\begin{widetext}
\begin{eqnarray}
&  & \sum_{\sigma_0}
e^{J_{10}\sigma_0\sigma_1+J_{20}\sigma_0\sigma_2+H_0\sigma_0+H_{ex}}[(J'_{10}\sigma_0\sigma_1+J'_{20}\sigma_0\sigma_2+H'_0\sigma_0+H'_{ex})^2
 + J''_{10}\sigma_0\sigma_1+J''_{20}\sigma_0\sigma_2+H''_0\sigma_0+H''_{ex}] \nonumber \\
 & =& e^{\delta F+J_{12}\sigma_1\sigma_2+\delta H_1 \sigma_1+\delta H_2 \sigma_2+H_{ex}} [(\delta
 F'+J'_{12}\sigma_1\sigma_2+\delta H'_1 \sigma_1+H'_2 \sigma_2+H'_{ex})^2  +\delta F''+J''_{12}\sigma_1\sigma_2+\delta H''_1 \sigma_1+H''_2 \sigma_2+H''_{ex}], \nonumber \\
\label{eq:series-in}
\end{eqnarray}
\end{widetext}
where the terms not involving $\sigma_0$ are collected in
$H'_{ex}(\sigma_1,\sigma_2,\cdots),H''_{ex}(\sigma_1,\sigma_2,\cdots)$.
In every step, we have $F\rightarrow F+\delta F$, $U\rightarrow
U+\delta U$ and $C\rightarrow C+\delta C$. Since
$H'_{ex},H''_{ex}$ contain the variables
$\sigma_1,\sigma_2,\cdots,U,C$, the coefficients before them in
the two sides of the above equation must be equal. Equating the
coefficients before $H'^2_{ex},H''_{ex}$ in both sides of the
above equation yields Eq. (\ref{eq:series-pa}), which is solved
above. Equating the coefficients before $H'_{ex}$ in both sides of
the above equation yields Eq. (\ref{eq:series-in1}), which is also
solved above. Equating the rest terms in the two sides  and using
Eqs. (\ref{eq:pa-2}) and Eqs. (\ref{eq:bp-u}) we get
\begin{widetext}
\begin{eqnarray}
\delta F''  +J''_{12}\pm \delta H''_1\pm \delta H''_2
& = & (H'_0\pm J'_{10}\pm J'_{20})^2(1-a^2_{\pm}) +(H''_0\pm J''_{10}\pm J''_{20})a_{\pm} \equiv z''_{\pm} \nonumber \\
\delta F'' -J''_{12}\pm \delta H''_1\mp \delta H''_2
& = & (H'_0\pm J'_{10}\mp J'_{20})^2(1-b^2_{\pm})+(H''_0\pm J''_{10}\mp J''_{20})b_{\pm} \equiv y''_{\pm}
\end{eqnarray}
\end{widetext}
The solution is given by
\begin{eqnarray}
\delta F'' & = & \frac{1}{4}(z''_{+}+z''_{-}+y''_++y''_-), \nonumber \\
\delta H''_1 & = & \frac{1}{2}(z''_+ +y''_+)-\delta F'' \nonumber \\
\delta H''_2 & = & \frac{1}{2}(z''_+ +y''_-)-\delta F'' \nonumber \\
J''_{12}  &  = & \frac{1}{2}(z''_++z''_-)-\delta F''.
\label{eq:bp-c}
\end{eqnarray}

In the algorithms, the transformations preserve these quantities
during every step. The accuracy is only limited  by the machines's
accuracy. With these algorithms, we can calculate the free energy,
internal energy and specific heat with the same accuracy. As
discussed in reference \cite{loh2}, the time of calculation is
proportional to $L^2 \times M$ if the lattice size is $M\times L$.
Then the accumulation of roundoff error is proportional to
$L\sqrt{M}$. Therefore the accuracy can reach $10^{-12}$ for a
lattice with size $100^2\times 100$ if all the variables are
assigned in the double precision format. Moreover the computing
time of the free energy, internal energy and specific heat on such
a large lattice is about one minute on an usual PC computer. It is
very efficient compared with other numerical method such as Monte
carlo simulation.

\section{Surface-Field-Site-Propagation algorithm for magnetization and correlation function}

The magnetization at site $j$ is defined by
\begin{equation}
<\sigma_j>=\sum_{\{\sigma_i \}}\sigma_j e^{-\beta
H(\{\sigma_i\})}/Z.
\label{eq:mag}
\end{equation}
To calculate this quantity for the two dimensional Ising model with a surface field, we need a new transformation, called Surface-Field-Site-Progation (SFSP) series reduction, which is  shown in Fig. 2a. The numerator can be calculated by SFSP algorithm. The denominator is the partition function, which can be calculated by BP algorithm. The spin $\sigma_j$ is called concerned spin. In Fig. 2c, we show a schematic of SFSP to calculate the numerator of the above equation with the concerned spin at the center, which is represented by a open square. The spins at the surface with external field are represented by open circles with a cross. The SP $Y-\Delta$ transformation and its inverse are proposed several years ago \cite{wu1}. SFSP series reduction, SP $Y-\Delta$ transformation and its inverse are the ingredients of SFSP algorithms.

\begin{figure}
 \begin{center}
    \resizebox{9cm}{7cm}{\includegraphics{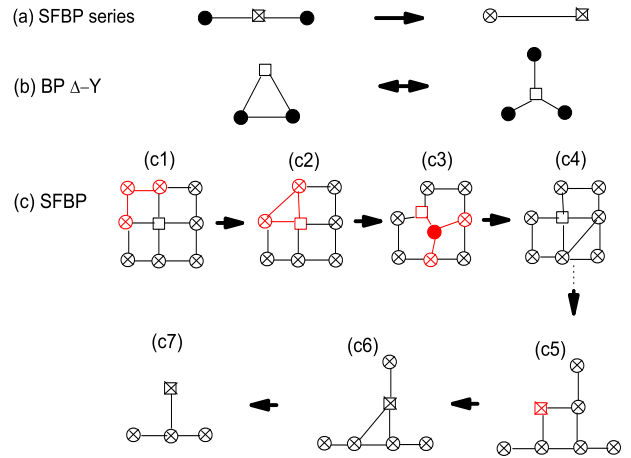}}
  \end{center}
\vskip -1cm
  \emph{
\caption{(Corlor online) The schematic of SFSP algorithm. (a) The SFSP seriese reduction transformation.
(b) The SP $\Delta-Y$ transformation and its inverse. (c) The schematic of SFBP on a $3\times 3$ lattice.
The concerned spin is at the center of lattice in (c1). Applying SFBP series reduction to the red part in (c1)
leads to (c2); Applying SP $\Delta-Y$ to the red part in (c2) leads to (c3); etc.  Applying SFSP series reduction to
the red part in (c5) leads to (c6). At last we get a small lattice in (c7), which can be calculated simply.}}
\end{figure}

SFSP series reduction corresponds to
integrating out a spin, on which a magnetic field is applied, with two neighbors, generating an effective
coupling $J_{12}$ and $\delta H_1,\delta H_2$ to make the
\begin{equation}
\sum_{\sigma_0}\sigma_0e^{\sigma_0(J_1\sigma_1+J_2\sigma_2+H_0)}\equiv
\sigma_1 e^{\delta F+J_{12}\sigma_1\sigma_2+\delta H_1 \sigma_1+\delta H_2 \sigma_2}, \label{eq:series}
\end{equation}
This equation should be valid for all $\sigma_1,\sigma_2$, then we get
\begin{eqnarray}
e^{\delta F+J_{12}\pm \delta H_1 \pm \delta H_2}=2\sinh(J_{10}+J_{20}\pm H_0)\equiv c_{\pm} \nonumber \\
e^{\delta F-J_{12}\pm \delta H_1 \mp \delta H_2}=2\sinh(J_{10}-J_{20}\pm H_0)\equiv d_{\pm} \nonumber \\
\end{eqnarray}
With variables $j_i=e^{-J_i}, \delta h_i=e^{-\delta H_i}$ and $\delta f=e^{\delta F}$, the solution of the
above equations is geiven by
\begin{eqnarray}
\delta f & = & ( c_+c_-d_+d_-)^{1/4},\hskip 0.5cm \delta h_1=\delta f/(c_+d_+)^{1/2},\nonumber \\
\delta h_2 & = & \delta f/(c_+d_-)^{1/2}, \hskip 0.5cm j_{12}=\delta f\delta h_1 \delta h_2/d_-
\end{eqnarray}

The schematic to calculate the correlation function is similar to that for the correlation function without surface field \cite{wu1}.
There are two concerned spins. As the concerned spins are shifted to the surface, one use the SFSP series reduction, rather than SP series reduction. Then one can get the correlation function.

For two surface spins, we have two methods to calculate their correlation. One method is using the SFSP algorithm discussed just above. Another one is using SFBP algorithm, discussed after Eq. (\ref{eq:bp-sp}). The former one can also be applied to two interior spins. The latter one can only be valid for the surfaces spins. However, the latter one has much high efficiency and accuracy than the former one to calculate the correlation between two surface spins. In the SFSP algorithm, the format of the variables must be complex, so it occupies more memory and the speed is slower. Moreover, in the SFSP algorithm, there are two parts (see Eq. (\ref{eq:mag}). The substraction of the logarithmic of partition function in the final result lower the accuracy greatly. For a lattice with size $100^2\times 100$, the accuracy in the SFSP algorithm can only reach $10^{-5}$ if the variables are in double complex format. If the SFBP algorithms is applied, the accuracy can be $10^{-11}$. Because the fluctuation near the surface is particularly important in the wetting transition, the SFBP algorithm for the correlation of two surface spins is quite useful.

\section{The Abraham's model for wetting transition}

As we know, the Ising model with surface field can be used to study the wetting transition.
We consider the Abraham's model \cite{abraham1980,parry1990,binder1995}
\begin{eqnarray}
-\beta \mathcal{H} & = & J\sum_{m=1}^{ M}[\sum_{l=1}^{L-1} \sigma(n,m)\sigma(n+1,m) \nonumber \\
& &+\sum_{n=1}^{N} \sigma(n,m)\sigma(n,m+1)] \nonumber \\
  & &+\sum_{m=1}^{M}[H_L\sigma(m,L)+H_1\sigma(m,1)]
\label{eq:lattice1}
\end{eqnarray}
where $ \beta=1/KT$, $J=\beta \tilde{J}$ and $H_i=\beta
\tilde{H}_i$.  We set $\tilde{J}/K=1$ in the numerical calculation throughout.

In this model, the surface field are applied only to
the top and the bottom layer. The field on the surfaces of
right and left sides are zero. This set-up is often referred to as
one with ¡°opposing boundaries¡± or ¡°competing walls¡± \cite{binder1995}. The
parallel size is $M$ and the perpendicular size is $L$.  We set
$\tilde{H}_L=-1$ throughout, which is equivalent to the fixed boundary condition
of setting the spins at $(L+1)$th layer be $-1$ if we add a layer
on the top. The bottom layer is the wall, where the field is positive $\tilde{H}_1>0$. In the following, we call the direction parallel to the wall the parallel direction and the one perpendicular to the wall the perpendicular direction.

In the thermodynamic limit, letting $M\rightarrow \infty$ followed by $L\rightarrow \infty$, a sharp surface phase transition occurs as a function of the control parameter $\tilde{H}_1$, assuming fixed $T < T_{cb}$, where $T_{cb}$ is the standard order-disorder critical temperature given by $T_{cb}=2/\ln(1+\sqrt{2})$ \cite{onsager}. For small $H_1$ with $H_1$ below the wetting point $H_w$ , the interface is localized at the bottom boundaries and this state is called partial wetting. For $H_1 > H_w$ it is (free-)energetically favorable for the interface to wander away
from the bottom boundaries and the system is in the complete wetting state.

The wetting transition is a singularity of the surface excess free energy $f_s (T,H,H_1)$, defined from standard decomposition of the
total free energy into the bulk term and boundary terms, for $ L \rightarrow \infty, M \rightarrow \infty$
\begin{equation}
F(T,H,H_1,L,M)/(LM)=f_b(T,H)+\frac{1}{L}f_s(T,H,H_1)
\end{equation}
where $H$ is bulk magnetic field. Our algorithm can not deal the case with nonzero bulk magnetic field.
With zero bluk magnetic field the surface excess free energy is expected to satisfy a scaling behavior
\begin{equation}
f_s\propto |t|^{2-\alpha_s}, \hskip 1cm t=1-T/T_w
\end{equation}
where $\alpha_s$ is the specific heat exponent for the wetting transition and $T_w$ is the wetting transition temperature for a given surface field $H_1$.  The excess specific heat should be
\begin{equation}
c_s=\beta^2\frac{\partial^2 f_s}{\partial \beta^2} \propto |t|^{-\alpha_s}
\label{eq:cs}
\end{equation}
If we fixed the temperature $T$ and change the surface field the scaling behavior is given by \cite{binder1983}
\begin{equation}
f_s\propto |H_1-H_w|^{2-\alpha_s}
\end{equation}
where $H_w$ is the transition point for the temperature $T$. Then we have the excess surface susceptibility
\begin{equation}
\chi^{(s)}_{11}=\frac{\partial ^2 f_s}{\partial H_1^2} \propto |H_1-H_w|^{\alpha_s}
\end{equation}
For the wetting transition, we have the scaling relation $2-\alpha_S=(d-1)\nu_{\parallel}$ \cite{albano2012}. The parallel correlation length exponent is $\nu_{\parallel}=2$, the perpendicular correlation length exponent is $\nu_{\perp}=1$, known from the Abraham's exact solution \cite{abraham1980}. Then we have $\alpha_s=0$ . The excess specific heat, surface magnetization has a finite jump at the transition point. In the following, we will study these critical behavior with our algorithm, and compare with the exact result. In all our
calculation, we set $M=L^2$ according to the anisotropic finite
size scaling \cite{albano2012}.

\subsection{Magnetization, specific heat and surface susceptibility}

 Using SFSP algorithm, one can calculate the magnetization of every spin. Because our boundary at the left and right sides are open. In the parallel direction, the magnetization is not homogeneous, especially near the boundary. In Fig. 3a, we show magnetization distribution for $1/T=0.5,\tilde{H}_1=\tilde{H}_w=0.4663995$ on the lattice with size $L=40,M=1600$. As one see, in a large center part of the system, the magnetization in the parallel direction is homogeneous. This is different from the periodic boundary condition, which is used in most study and there is no such a inhomogeneity. This inhomogeneity will cause some difference between our results and those with periodic boundary condition. However, this inhomogeneity will become weaker as the system size increases. Moreover, the calculation on very large size lattices, say $L=120, M=L^2$ can be carried out easily with our algorithm. Therefore it does cause serious problem in the study.

\begin{figure}
 \begin{center}
    \resizebox{9cm}{7cm}{\includegraphics{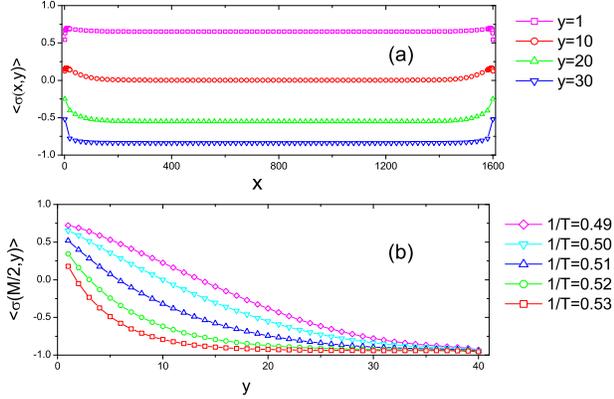}}
  \end{center}
\vskip -1cm
  \emph{
\caption{(Corlor online) Some typical magnetization profile on the lattice with size $L=40, M=1600$. (a) The magnetization of the spins at first, $10$th, $20$th, $30$th row (along the parallel direction) with the temperature is $T=2.0$ and $\tilde{H}_1=\tilde{H}_w=0.4663995\cdots$. (b) The magnetization profile of spins at the middle column along the perpendicular direction with five different temperature and $\tilde{H}_1=0.4663995$. }}
\end{figure}

Figure 3b shows the magnetization profile of the middle column spins at five different temperatures. As one can expected, the magnetization profiles near the right and left sides are different form that of the middle column.

Figure 4 shows the magnetization defined in the following approximate way
\begin{equation}
m=\frac{1}{L}\sum_{l=1}^{L}<\sigma(M/2,l)>.
\end{equation}
Recall $M$ is the parallel size and $L$ is the perpendicular size. In another words, we take the average of the magnetization of  middle column spins as the average of the whole system approximately. Because we keep $M=L^2$, the inhomogeneity in the parallel direction does not cause significant effect. Figure 4 plots $m$ vs $T$ for three choices of $L$ for $M=L^2$, while the inset replots the data in scaled form.

\begin{figure}
 \begin{center}
    \resizebox{9cm}{7cm}{\includegraphics{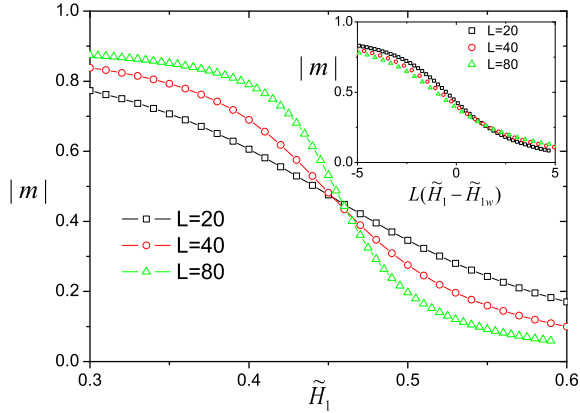}}
  \end{center}
\vskip -1cm
  \emph{
\caption{(Corlor online) The intersections of the magnetization of different sizes. The inset show the scaling plot of $|m|$ vs $L(\tilde{H}_1-\tilde{H}_w)$.The inverse of temperature $T=2.0$. The exact transition point is $\tilde{H}_w=0.4663995\cdots$.}}
\end{figure}

In our calculation the excessive specific heat is defined by
\begin{equation}
c_S(T,H_1,L)=\frac{1}{M}[C(T,H_1,L)-C(T,0,L)]
\end{equation}
where $C(T,H_1,L)$ is the specific heat with surface $H_1$ and $C(T,0,L)$ is the specific heat with zero surface field at the bottom layer (the wall). For $M,L\rightarrow \infty$ this quantity is an alternative definition in Eq. (\ref{eq:cs}).

\begin{equation}
\chi^{(s)}_{11}=\chi_{11}(T,H_1,L)-\chi_{11}(T,0,L)
\end{equation}
where $\chi (T,H_1,L)$ is the specific heat with surface $H_1$ and $\chi (T,0,L)$ is the specific heat with zero surface field at the bottom layer. They are calculated by the SFBP algorithm mentioned in section 2.

\begin{figure}
 \begin{center}
    \resizebox{9cm}{7cm}{\includegraphics{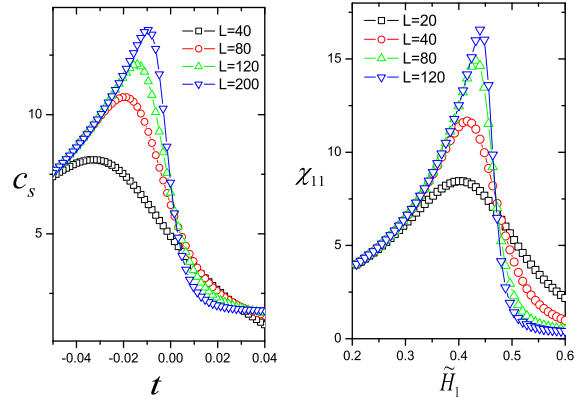}}
  \end{center}
\vskip -1cm
  \emph{
\caption{(Corlor online) (a) The excess specific heat of different sizes.  (b) Excess surface susceptibility of different sizes. For both cases the temperature is $T=2.0$ }}
\end{figure}

Because $\alpha_s=0$, there is a finite jump in the surface excess specific heat and the surface susceptibility. This is clearly shown in the Fig. 5. As the size of the system increases drastic jump develops.

\subsection{Locating the transition point}

The intersections of the order parameters of different size coincide approximately, as shown in Fig. 4.  As the system  sizes approach infinity, the intersections should be coincide exactly. and the  intersecting point is the transition point. Therefore we can obtain the transition point through studying the convergence of the intersections. We calculate the intersections of $L=20$ and $L=21$; $L=25$ and $L=26$; $\cdots$;  $L=85$ and $L=86$, then  fit these data to extrapolate the intersection of  $L=\infty$ and $L=\infty+1$ , which should be the transition point. Due the high accuracy, we can determine these intersections to $10^{-6}$ in double precision format. Then we fit the data with the formula $\tilde{H}(L)=\tilde{H}_w+\sum_{k=1}^{k_{max}}A_k(L+0.5)^{-k}$, where $\tilde{H}(L)$ is surface field at the intersection of size $L$ and $L+1$. Because the intersection belongs to the size of $L$ and $L+1$, we set the size in the expansion to be $L+0.5$. We show some fitting results in the Table I, which are labelled by $\tilde{H}_w^{(mag)}$. In the fitting, we use $k_{max}=4$.

\begin{table}
 \caption{ The comparison of wetting transition temperature obtained numerically with the exact result given by Eq. (\ref{eq:TCR})  }
\begin{tabular}{llll}

\hline
$\beta $       & $\tilde{H}_w^{(ex)}$        &$\tilde{H}_w^{(mag)} $ & $\tilde{H}_w^{(sus)}$   \\
$0.48$         & $0.389150745\cdots$ & $0.38923(5) $ & $0.389155(2) $  \\
$0.50$         & $0.466395503\cdots$ & $0.46634(5) $ & $ 0.466396(1) $ \\
$0.52$         & $0.526678446\cdots$ & $0.52673(5) $ & $0.5266786(4) $ \\
$0.60$         & $0.683832908\cdots$ & $0.68384(4) $ & $0.683832910(4)$ \\
\hline
\end{tabular}
\label{fitfen1}
\end{table}

Our estimates for the transition point agree with the exact result of Abraham \cite{abraham1980},
\begin{equation}
e^{2\beta}[\cosh 2\beta -\cosh 2\tilde{H}_w \beta]=\sinh 2\beta.
\label{eq:TCR}
\end{equation}
In table I, the exact results are labelled by $\tilde{H}_w^{(ex)}$. The results of $\tilde{H}_w^{(sus)}$ are obtained by another method introduced below.

As we can see in Fig. 4b, the surface susceptibilities of different size intersect at the same point approximately. This provides us another way to locate the transition point. According to the same route, we can locate the transition through the convergence of the surface susceptibility. See Fig. 3b for the intersections of the surface susceptibility $\chi_{11}$. The convergence of the surface susceptibility also gives the transition point. We calculate the intersections of $L$ and $L+1$.  We calculate the intersections of L=20 and L=21; L=25 and L=26; ¡­.  etc., then  fit these data to extrapolate the intersection of  $L=\infty$ and $L=\infty+1$ , which should be the transition point. This method is much more accurate than the above one. The results with this methods are labelled by $\tilde{H}_w^{(sus)}$ in table I. As one can see, the accuracy reaches to $10^{-8}$ for $\beta=0.6$. Moreover, this method is more efficient the above one. It costs much less computing time. The real format is used this way and the complex format is used in the way of magnetization. In addition, we have to calculate the magnetization of $L$ spins.

\subsection{Correlation function and correlation length exponents}

\begin{figure}
 \begin{center}
    \resizebox{9cm}{7cm}{\includegraphics{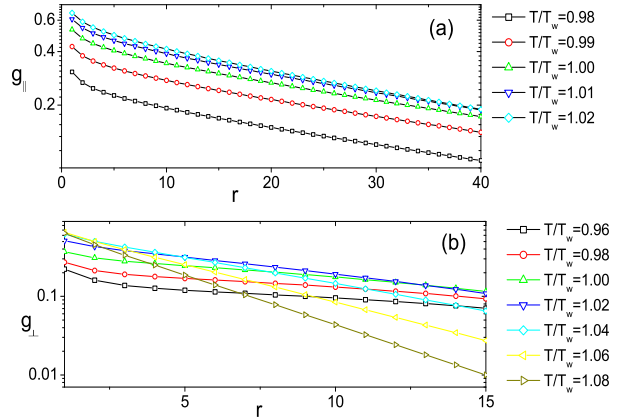}}
  \end{center}
\vskip -1cm
  \emph{
\caption{(Corlor online) Typical parallel correlation functions on the lattice with $L=40,M=L^2$. (a) The parallel correlation function $g_{\parallel}$ is calculated for two spins of which position are $(M/2,L/2)$ and $(M/2+r,L/2)$. (b) The perpendicular correlation function $g_{\perp}$ is calculated for two spins of which position are $(M/2,1)$ and $(M/2,r+1)$ at the middle column.}}
\end{figure}

Using SFSP algorithm we can calculate the correlation function $
g({\textbf r}_j,{\textbf r}_k)=\sum_{\{\sigma_i\}}\sigma_j \sigma_k e^{-\beta H}/Z-<\sigma_j><\sigma_k>
$. Figure 6 show some typical correlation functions for surface field $\tilde{H}_{1}=0.4663995$, at which the wetting transition temperature $T_w=2$. In the figure 6(a) the correlation functions for five different temperatures are shown for the parallel direction. In the figure 6(b), the correlation functions for the perpendicular direction are shown. As one can see, the correlation function at large distance decays with distance exponentially, i.e. $g(r)\approx c_0 \exp(-r/ \xi)$ at large distance $r$. The inverse of slope of  $d \ln g(r)/dr$ at large distance is the correlation length.

\begin{figure}
 \begin{center}
    \resizebox{9cm}{7cm}{\includegraphics{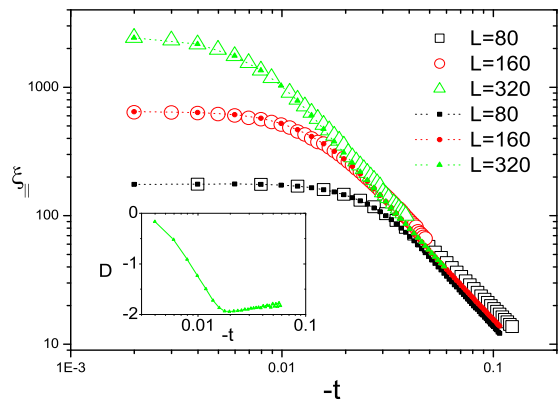}}
  \end{center}
\vskip -1cm
  \emph{
\caption{(Corlor online) Parallel correlation length near the transition point for different size. $D=d\ln(\xi_{\parallel})/d\ln(-t)$ is the slope of the tangent.}}
\end{figure}

Near the transition point, the parallel correlation length obey the scaling law
\begin{equation}
\xi_{\parallel}\propto |t|^{\nu_{\parallel}}
\end{equation}
where the Abraham's  exact result gives the exponent $\nu_{\parallel}=2.0$  \cite{abraham1980}. To verify this result, we calculate
the correlation length for $L=80,160,320$ and $M=L^2$. The result is shown in Fig. 7. The data in scatter is for the correlation
between two spins at $(M/2+r/2,L/2)$ and $(M/2-r/2,L/2)$, at the middle row of the system. They are obtained by SFSP algorithm in
complex*32 format. The data in solid line is for the correlation between two spins at $(M/2+r/2,1)$ and $(M/2-r/2,1)$, where is
bottom surface. They are obtained by SFBP algorithm in real quadruple precision format. As one can see the two results are
approximately the same. The difference for these two results are the correlation amplitude. Near the transition point, $|t|$ is small, the magnetization at the middle row is already large while the magnetization at the wall (the first layer spins) is small. Therefore the fluctuation amplitude at the middle row is smaller than at the wall.

For $L=320$, in the temperature interval $0.01<|t|<0.1$, it has $\xi_{\parallel}\propto |t|^{-1.88}$. In other words, the average slope of the curve $\log \xi_{\parallel}$ vs $\log (-t)$ is $-1.88$. The inset shows the slope of the tangent of the curve. The  absolute value of the slope increases as $|t|$ decreases first. Its maximum value is $1.95$. At this point, we may say that the effective exponent is $\nu_{\parallel}=-1.95$. It decreases as the $|t|$ decreases further due to the finite size effect. Therefore, our best estimate of the parallel correlation length exponent is $1.95$.

\begin{figure}
 \begin{center}
    \resizebox{9cm}{7cm}{\includegraphics{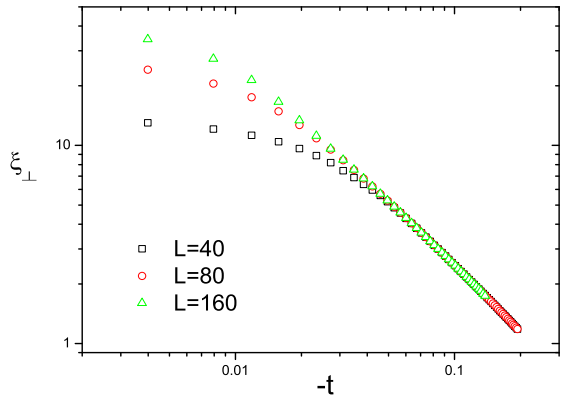}}
  \end{center}
\vskip -1cm
  \emph{
\caption{(Corlor online) Perpendicular correlation length near the transition point for different size. }}
\end{figure}

Figure 8 shows the result on the perpendicular correlation length. Near the transition point, the perpendicular correlation length obey the scaling law
\begin{equation}
\xi_{\perp}\propto |t| ^{ -\nu_{\perp}}
\end{equation}
where the exponent is $\nu_{\perp}=1$ in the exact result of Abraham \cite{abraham1980}. To verify this result, we calculate the correlation length for $L=40,80,160$ and $M=L^2$. The result is shown in Fig. 6b.  The data are calculated for the correlation between two spins at $(M/2+r/2,1)$ and $(M/2-r/2,1)$, where is bottom surface. For $L=160$, in the temperature interval $0.01<|t|<0.1$, it has $\xi_{\perp}\propto |t|^{-1.04}$. Therefore our best estimate of perpendicular correlation length exponent is $1.04$.

\section{Summary and discussion}

We have developed a set of efficient algorithms to study the two dimensional Ising model with surface field, which can be used to study the wetting transition. With these algorithms, we can calculate the magnetization, specific heat, surface susceptibility, correlation function, etc. very accurately. They are also highly efficient and can be applied on lattices with very large size. These algorithms provide us another powerful weapon to cope with the wetting transition.

Extending Abraham's model the Ising model with a surface field has been extensively and intensively studied. Lipowski showed that the corner wetting transition temperature is different from the flat wall \cite{lipowski}. Late on there has been a lot study on this corner filling  transition \cite{abraham2002}. Forgacs et. al showed that introduction of a line defect far from the wall can turn the wetting transition to be first order \cite{forgacs}. This is an interesting model because it has a first order transition and can be solved exactly. The effect of nonuniform surface field is also studied recently \cite{albano2012a}. Our algorithms can be applied to these problems.

These algorithms can also be applied to the disordered systems, which are very difficult to cope with. Because in our algorithms the bonds and surface field can be random, the wetting transition with bond randomness and surface field randomness can be studied directly.  To our knowledge, the numerical work on this field is rare. The high efficiency, accuracy may make this method advantageous to other numerical ones.

The author thanks J. O. Indekeu for useful discussions.

\end{document}